\documentclass[aps,prl,twocolumn,showpacs,superscriptaddress,groupedaddress]{revtex4-1}  
\usepackage{graphicx}
\usepackage{dcolumn}
\usepackage{bm}
\usepackage{amssymb}
\usepackage{color}
\usepackage{multirow}
\usepackage{epsfig}
\usepackage{array}
\usepackage{amssymb}
\usepackage{amsfonts}
\usepackage{amsmath}
\usepackage{times}

\begin{document}

\title{Robust Reconstruction of Complex Networks from Sparse Data}

\author{Xiao Han}
\affiliation{School of Systems Science, Beijing Normal University,
Beijing, 100875, P. R. China}

\author{Zhesi Shen}
\affiliation{School of Systems Science, Beijing Normal University,
Beijing, 100875, P. R. China}

\author{Wen-Xu Wang}\email{wenxuwang@bnu.edu.cn}
\affiliation{School of Systems Science, Beijing Normal University,
Beijing, 100875, P. R. China}

\author{Zengru Di}
\affiliation{School of Systems Science, Beijing Normal University,
Beijing, 100875, P. R. China}

\date{\today}


\begin{abstract}
Reconstructing complex networks from measurable data is a fundamental problem for understanding and controlling
collective dynamics of complex networked systems. However, a significant challenge arises
when we attempt to decode structural information hidden in limited amounts of data accompanied by noise and
in the presence of inaccessible nodes. Here, we develop a general framework for robust reconstruction
of complex networks from sparse and noisy data. Specifically, we decompose the task of reconstructing the
whole network into recovering local structures centered at each node. Thus, the natural sparsity
of complex networks ensures a conversion from the local structure reconstruction into a sparse
signal reconstruction problem that can be addressed by using the lasso,
a convex optimization method. We apply our method
to evolutionary games, transportation and communication processes taking place in a variety of
model and real complex networks, finding that universal high reconstruction accuracy can be
achieved from sparse data in spite of noise in time series and missing data of partial nodes.
Our approach opens new routes to the network reconstruction problem and has potential applications
in a wide range of fields.
\end{abstract}


\maketitle

Complex networked systems are common in many fields~\cite{taming,boccaletti,newman}. The need to
ascertain collective dynamics of such systems to control them is shared among
different scientific communities~\cite{hecker,timmereview,credit}. Much evidence has demonstrated that interaction
patterns among dynamical elements captured by a complex network play deterministic roles in
collective dynamics~\cite{strogatz}. It is thus imperative to study a complex networked system as a whole
rather than study each component separately to offer a comprehensive understanding
of the whole system~\cite{takeover}. However, we are often incapable of directly accessing network
structures; instead, only limited observable data are available~\cite{raredata}, raising the need for
network reconstruction approaches to uncovering network structures from data.
Network reconstruction, the inverse problem, is challenging because structural information
is hidden in measurable data in an unknown manner and the solution space of all possible structural
configurations is of extremely high dimension. So far a number of approaches have been proposed to address
the inverse problem~\cite{hecker,timmereview,raredata,connectivity,topology,catastrophe,pg,shen,auroc,indirect,
feizi}. However, accurate and robust reconstruction of large complex networks is still a challenging problem,
especially given limited measurements disturbed by noise and unexpected factors.


In this letter, we develop a general framework to reconcile the contradiction between
the robustness of reconstructing complex networks and limits on our ability to
access sufficient amounts of data required by conventional approaches.
The key lies in converting the network reconstruction problem into a sparse signal reconstruction
problem that can be addressed by exploiting the lasso, a convex optimization algorithm~\cite{hastie,python}.
In particular, reconstructing the whole network structure can be achieved by inferring local connections
of each node individually via our framework. The natural sparsity of complex networks
suggests that on average the number of real connections of a node is much less than
the number of all possible connections, i.e., the size of a network. Thus, to identify direct
neighbors of a node from the pool of all nodes in a network is analogous to
the problem of sparse signal reconstruction. By using the lasso that incorporates
both an error control term and an L1-norm, the neighbors of each node can be reliably identified
from a small amount of data that can be much less than the size of a network. The L1-norm, according to
the compressed sensing theory~\cite{CS}, ensures the sparse data requirement while, simultaneously, the error control term
ensures the robustness of reconstruction against noise and missing nodes.
The whole network can then be assembled by simply matching neighboring sets of all nodes.
We will validate our reconstruction framework by considering three representative
dynamics, including ultimatum games~\cite{altruism}, transportation~\cite{transportation} and communications~\cite{communication},
taking place in both homogeneous and heterogeneous networks. Our approach opens
new routes towards understanding and controlling complex networked systems and has
implications for many social, technical and biological networks.


We articulate our reconstruction framework by taking ultimatum games as a
representative example. We then apply the framework to the transportation of electrical current and
communications via sending data packets.

In evolutionary ultimatum games (UG) on networks, each node is occupied by a player.
In each round, player $i$ plays the UG twice with each of his/her neighbors, both as a proposer
and a responder with strategy ($p_i$, $q_i$), where $p_i$ denotes the amount offered to the other
player if $i$ proposes and $q_i$ denotes the minimum acceptance level if $i$ responds~\cite{nowak,szolnoki}.
The profit of player $i$ obtained in the game with player $j$ is calculated as follows
\begin{eqnarray}
U_{ij}=\left\{\begin{array}{cc}
                   p_j+1-p_i & p_i\geq q_j\ \hbox{and}\ p_j\geq q_i\\
                   1-p_i     & p_i\geq q_j\ \hbox{and}\ p_j<q_i\\
                   p_j       & p_i< q_j\ \hbox{and}\ p_j\geq q_i\\
                   0         & p_i< q_j\ \hbox{and}\ p_j< q_i\\
                  \end{array}\right.
\label{eq:UG}
\end{eqnarray}
where $p_i,p_j\in[0,1]$. The payoff $g_i$ of $i$ at a round is the sum of all profits from playing UG with
$i$'s neighbors, i.e., $g_i=\sum_{j\in\Gamma_i}U_{ij}$, where $\Gamma_i$ denotes the set of $i$'s neighbors.
In each round, all participants play the UG with their direct neighbors simultaneously and gain payoffs.
Players update their strategies ($p,q$) in each round by learning from one of their neighbors with the highest payoffs.
To be concrete, player $i$ selects the neighbor with
the maximum payoff $g_{\max}(t)$ and takes over the neighbor's strategy with probability
$W(i\leftarrow {\max})=g_{\max}(t)/[g_i(t)+\sum_{j\in\Gamma_i}g_j(t)]$~\cite{game:review}.
To better mimic real situations, random mutation rates
are included in each round: all players adjust their ($p,q$) according to
$(p_i(t+1), q_i(t+1))=(p_i(t)+\delta, q_i(t)+\delta)$,
where $\delta\in[-\varepsilon,\varepsilon]$ is a small random number~\cite{kuperman}.
Without loss of generality, we set $\varepsilon=0.05$ and $p,q\in[0,1]$.
During the evolution of UG, we assume that only the time series of
($p_i(t),q_i(t)$) and $g_i(t)$ ($i=1,\cdots,N$) are measurable.

The network reconstruction can be initiated from the
relationship between strategies $(p_i(t),q_i(t))$ and payoffs $g_i(t)$.
Note that $g_i(t) = \sum_{j=1,j\neq i}^N a_{ij}U_{ij}$, where
$a_{ij}=1$ if player $i$ and $j$ are connected and $a_{ij}=0$ otherwise.
Moreover, $U_{ij}$
is exclusively determined by the strategies of $i$ and $j$. These imply that
hidden interactions between $i$ and its neighbors can be extracted from the relationship between strategies
and payoffs, enabling the inference of $i$'s links
based solely on the strategies and payoffs.
Necessary information for recovering $i$'s links can be acquired
with respect to different time $t$. Specifically,
for $M$ accessible time instances $t_1,\cdots , t_M$,
we convert the reconstruction problem into the matrix form $\mathbf{Y}_i=\Phi_i\times \mathbf{X}_i$:
\begin{equation}
\left[\begin{array}{c}
  y_i(t_1) \\
  y_i(t_2) \\
  \vdots\\
  y_i(t_M)
\end{array}\right]=
\left[\begin{array}{cccc}
               \phi_{i1}(t_1) & \phi_{i2}(t_1) & ... & \phi_{iN}(t_1)\\
               \phi_{i1}(t_2) & \phi_{i2}(t_2) & ... & \phi_{iN}(t_2)\\
               \vdots & \vdots & \vdots & \vdots \\
               \phi_{i1}(t_M) & \phi_{i2}(t_M) & ... & \phi_{iN}(t_M)\\
              \end{array}\right]
\left[
  \begin{array}{c}
    x_{i1} \\
    x_{i2} \\
    \vdots \\
    x_{iN} \\
  \end{array}
\right],
\end{equation}
where $\mathbf{Y}_i \in \mathbb{R}^{M\times 1}$ is the payoff vector of $i$ with
$y_i(t_\mu)=g_i(t_\mu)$ $(\mu = 1,\cdots, M)$,
$\mathbf{X}_i \in \mathbb{R}^{N\times 1}$ is the neighboring vector of $i$ with
$x_{ij}=a_{ij}$ $(j=1,\cdots, N)$
and $\Phi_i \in \mathbb{R}^{M\times N}$ is the virtual-payoff matrix of $i$ with
$\phi_{ij}(t_\mu)=U_{ij}(t_\mu)$.

%

\begin{figure}
\begin{center}
\epsfig{figure=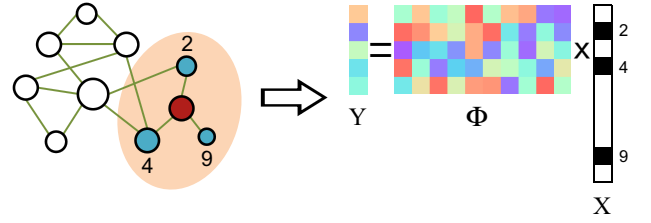,width=\linewidth}
\caption{Illustration of reconstructing the local structure of a node.
For the red node with three neighbors, $\#$2, $\#$4 and $\#$9 in blue, we can
establish vector $\mathbf{Y}$ and matrix $\Phi$ in the reconstruction
form $\mathbf{Y}=\Phi \mathbf{X}$ from data, where vector $\mathbf{X}$ captures
the neighbors of the red node. If the reconstruction is accurate,
elements in the 2nd, 4th and 9th rows of $\mathbf{X}$ corresponding to nodes
$\#$2, $\#$4 and $\#$9 will be nonzero values (black color),
while the other elements are zero (white color). The length of
$\mathbf{X}$ is $N$, which is in general much larger than the average degree
of a node, say, three neighbors, assuring the sparsity of $\mathbf{X}$.
In a similar fashion, the local structure of each node can be recovered
from relatively small amounts of data compared to the network size by
using the lasso. Note that only one set of data is used to reconstruct
local structures of different nodes, which ensures the sparse data requirement.
}
\label{fig:illustration}
\end{center}
\end{figure}

Because $U_{ij}(t)$ is determined by $(p_i(t),q_i(t))$ and
$(p_j(t),q_j(t))$ according to Eq.~(\ref{eq:UG}), $\mathbf{Y}_i$ and $\Phi_i$ can be
collected or calculated directly from the time series of strategies and payoffs.
Our goal is to reconstruct $\mathbf{X}_i$ from $\mathbf{Y}_i$ and $\Phi_i$.
Note that the number of nonzero elements in $\mathbf{X}_i$, i.e., the number of the neighbors of $i$,
is usually much less than length $N$ of $\mathbf{X}_i$. This indicates that
$\mathbf{X}_i$ is sparse, which is ensured by the natural sparsity of complex networks. An intuitive illustration of the
reconstruction method is shown in Fig.~\ref{fig:illustration}. Thus, the problem of identifying the neighborhood
of $i$ is transformed into that of sparse signal reconstruction, which can be addressed
by using the lasso.

The lasso is a convex optimization method for solving
\begin{equation}
\min_{\mathbf{X}_i}\left \{\frac{1}{2M}\|\mathbf{Y}_{i}-\Phi_{i}\mathbf{X}_i\|_{2}^2+\lambda\|\mathbf{X}_{i}\|_1 \right\},
\end{equation}
where $\lambda$ is a nonnegative regularization parameter~\cite{hastie,python}. The sparsity of the solution is
ensured by $\|\mathbf{X}_{i}\|_1$ in the lasso according to the compressed sensing theory~\cite{CS}.
Meanwhile, the least square term $\|\mathbf{Y}_{i}-\Phi_{i}\mathbf{X}_i\|_{2}^2$ makes
the solution more robust against noise in time series and missing data
of partial nodes than would the $L_1$-norm-based optimization method.

The neighborhood of $i$ is given by the reconstructed vector $\mathbf{X}_i$,
in which all nonzero elements correspond to direct neighbors of $i$. In a similar fashion,
we construct the reconstruction equations of all nodes, yielding the neighboring sets of
all nodes. The whole network can then be assembled by simply matching the neighborhoods
of nodes. Due to the sparsity of $\mathbf{X}_i$, it can be reconstructed by using
the lasso from a small amount of data that are much less than the length of $\mathbf{X}_i$,
i.e., network size $N$. Although we infer the local structure of each node separately by
constructing its own reconstruction equation, we only use one set of data sampling in
time series. This enables a sparse data requirement for recovering the whole network.



We consider current transportation in a network consisting of
resistors~\cite{transportation}. The resistance of a resistor between node $i$ and $j$ is denoted by
$r_{ij}$. If $i$ and $j$ are not directly connected by a resistor, $r_{ij}=\infty$.
For arbitrary node $i$, according to Kirchhoff's law, we have
\begin{equation}
\sum_{j=1}^{N}\frac{a_{ij}}{r_{ij}}(V_i-V_j)=I_i,
\label{eq:Kirchhoff}
\end{equation}
where $V_i$ and $V_j$ are the voltage at $i$ and $j$ and $I_i$ is the total electrical
current at $i$. To better mimic real power networks, alternating current is considered.
Specifically, at node $i$, $V_i=\bar{V}\sin[(\omega+\Delta\omega_i)t]$, where the constant
$\bar{V}$ is the voltage peak, $\omega$ is frequency and $\Delta\omega_i$ is perturbation.
Without loss of generality, we set $\bar{V}=1$, $\omega =10^3$ and the random number
$\Delta\omega_i \in [0,20]$. Given voltages at nodes and resistances of links, currents
at nodes can be calculated according to Kirchhoff's laws at different time constants.
We assume that only voltages and electrical currents at nodes are measurable and our purpose is
to reconstruct the resistor network. In an analogy with networked ultimatum games, based on
Eq.~(\ref{eq:Kirchhoff}), we can establish the reconstruction equation $\mathbf{Y}_i=\Phi_i\times \mathbf{X}_i$
with respect to time constants $t_1,\cdots,t_M$, where $y_i(t_\mu)=I_i(t_\mu)$,
$x_{ij}=1/r_{ij}$ and $\phi_{ij}(t_\mu)=V_i(t_\mu)-V_j(t_\mu)$ with
$\mu =1,\cdots,M$ and $j=1,\cdots,N$. Here, if $i$ and $j$ are connected by a resistor,
$x_{ij}=1/r_{ij}$ is nonzero; otherwise, $x_{ij}=0$. Thus, the neighboring vector $\mathbf{X}_i$ is sparse and can
be reconstructed by using the lasso from a small amount of data. Analogously, the whole
network can be recovered by separately reconstructing the neighboring vectors of all nodes.

\begin{figure}
\begin{center}
\epsfig{figure=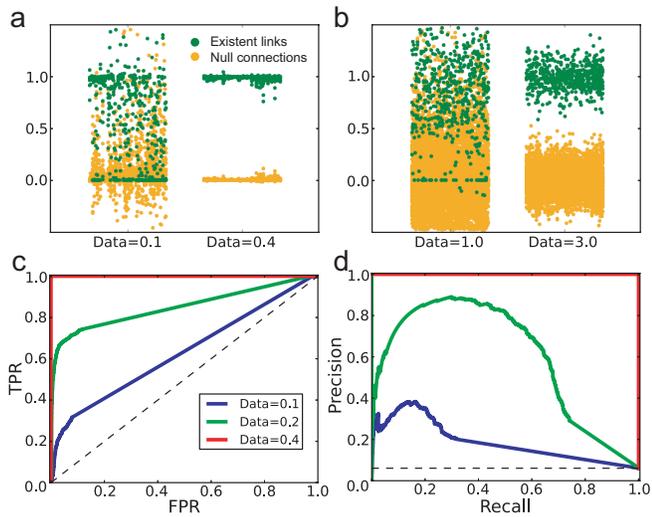,width=\linewidth}
\caption{Reconstructed values of elements in vector $\mathbf{X}$ for UG on small-world networks~\cite{ws}
for different data amounts (a) without measurement noise and (b) with Gaussian noise ($\mathcal{N}(0,0.3^2)$).
(c) TPR versus FPR and (d) Precision versus Recall for different data amounts for
UG on WS small-world networks without noise. In (c) and (d), the dashed lines represent the results of completely
random guesses. The network size $N$ is 100, and the average degree $\langle k\rangle=6$. Rewiring probability of
small-world networks is 0.3. There are no externally inaccessible nodes. The parameter $\lambda$ is set to
be $10^{-3}$. We have tested a wide range value of $\lambda$, finding that
optimal reconstruction performance can be achieved in range $[10^{-4},10^{-2}]$ and the reconstruction
performance in the range is insensitive to $\lambda$. Thus, we set $\lambda =10^{-3}$
for all reconstructions.
}
\label{fig:TprFpr}
\end{center}
\end{figure}

\begin{table*}
\caption{Minimum data for achieving at least 0.95 AUROC and AUPR simultaneously for three types of
dynamics, UG, current transportation and communications in combination with three types of networks,
random (ER), small-world (SW) and scale-free (SF). Here, $N$ is
network size, $\langle k\rangle$ is average degree, $\sigma$ is the variance of Gaussian noise,
and $n_\text{m}$ is the proportion of externally inaccessible nodes whose data are missing. Data
denote the amount of data divided by network size. The
results are obtained by averaging over 10 independent realizations. RN denotes resistor network, and
CN denotes communication network. More details of the reconstruction
performance as a function of data amount for different cases can be found in~\cite{SM}.}
\begin{tabular}{p{1.3cm}<{\centering}p{1.3cm}<{\centering}p{1.3cm}<{\centering}p{1.3cm}<{\centering}
p{3.8cm}<{\centering}p{3.8cm}<{\centering}p{3.8cm}<{\centering}}
\hline
\hline
\rule{0pt}{3ex}
 \multirow{1}{*}{$N$} & \multirow{1}{*}{$\langle k\rangle$} & \multirow{1}{*}{$\sigma$} & $n_m$ & UG & RN & CN \\
 & & & & \multicolumn{3}{c}{(ER\ /\ SW\ /\ SF)}\\[0.5ex]
\hline
\rule{0pt}{3ex} \multirow{7}{*}{100} & 6 & 0 & 0 & 0.38\ /\ 0.36\ /\ 0.41 & 0.28\ /\ 0.25\ /\ 0.32 & 0.30\ /\ 0.28\ /\ 0.30\\[0.3ex]
  & 6 & 0.05 & 0 & 0.44\ /\ 0.43\ /\ 0.47 & 0.29\ /\ 0.26\ /\ 0.37 & 0.34\ /\ 0.31\ /\ 0.34\\[0.3ex]
  & 6 & 0.3 & 0 & 1.68\ /\ 1.75\ /\ 1.60 & 0.32\ /\ 0.29\ /\ 0.38 & 1.72\ /\ 1.81\ /\ 1.80 \\[0.3ex]
  & 6 & 0 & 0.05 & 0.61\ /\ 0.55\ /\ 0.64 & 1.61\ /\ 1.65\ /\ 1.60 & 1.33\ /\ 1.19\ /\ 1.32 \\[0.3ex]
  & 6 & 0 & 0.3 & 2.33\ /\ 2.03\ /\ 2.14 & 5.74\ /\ 8.51\ /\ 8.50 & 5.38\ /\ 6.23\ /\ 6.20 \\[0.3ex]
  & 12 & 0 & 0 & 0.46\ /\ 0.47\ /\ 0.52 & 0.37\ /\ 0\ /\ 35\ /\ 0.42 & 0.42\ /\ 0.40\ /\ 0.42\\[0.3ex]
  & 18 & 0 & 0 & 0.53\ /\ 0.53\ /\ 0.58 & 0.44\ /\ 0.44\ /\ 0.50 & 0.50\ /\ 0.50\ /\ 0.50 \\[0.3ex]
  500 & 6 & 0 & 0 & 0.120\ /\ 0.116\ /\ 0.132 & 0.094\ /\ 0.080\ /\ 0.120 & 0.094\ /\ 0.088\ /\ 0.100\\[0.3ex]
  1000 & 6 & 0 & 0 & 0.071\ /\ 0.068\ /\ 0.078 & 0.058\ /\ 0.049\ /\ 0.079 & 0.055\ /\ 0.050\ /\ 0.055\\[0.5ex]
\hline
\hline
\end{tabular}
\label{tab:simu}
\end{table*}

We propose a simple network model to capture communications in populations
via phones, emails, etc. At each time, individual $i$ may contact one of his/her neighbors
$j$ according to probability $w_{ij}$ by sending data packets. If $i$ and $j$ are not connected,
$w_{ij}=0$. In a period, the total incoming flux $f_i$ of $i$ can be described as
\begin{equation}
f_{i}= \sum_{j=1}^{N}w_{ij}\widetilde{f}_{j},
\label{eq:flux}
\end{equation}
where $\widetilde{f}_{j}$ is the total outgoing flux from $j$ to its neighbors in the period
and $\sum_{j=1}^{N}w_{ij}=1$. Equation~(\ref{eq:flux}) is valid because of the flux conservation in
the network. In the real situation, $\widetilde{f}_{j}$ usually fluctuates with time, providing an
independent relationship between incoming and outgoing fluxes for constructing the reconstruction
equation $\mathbf{Y}_i=\Phi_i\times \mathbf{X}_i$. Here, $y_i(t_\mu)=f_{i}(t_\mu)$ is the total
incoming flux of $i$ at time period $t_\mu$, $\phi_{ij}(t_\mu)=\widetilde{f}_{j}(t_\mu)$ is the
total outgoing flux of $j$ at time period $t_\mu$, and $x_{ij}=w_{ij}$ captures connections between
$i$ and its neighbors. Given the total incoming and outgoing fluxes of nodes that can be measured without
the need of any network information and communication content, we can as well use the lasso
to reconstruct the neighboring set of node $i$ and those of the other nodes, such that
full reconstruction of the whole network is achieved from sparse data.

We simulate ultimatum games, electrical currents and communications on both
homogeneous and heterogeneous networks, including random~\cite{er}, small-world~\cite{ws}
and scale-free~\cite{ba} networks. For the three types of dynamical processes, we record
strategies and payoffs of players, voltages and currents, and incoming
and outgoing fluxes at nodes at different times, to apply our
reconstruction method with respect to different amounts of Data (Data$\equiv M/N$,
where $M$ is the number of accessible time instances in the
time series). Figure~\ref{fig:TprFpr} shows the results of networked ultimatum games.
For very small amounts of data, e.g., Data=0.1, links are difficult identify
because of the mixture of reconstructed elements in $\mathbf{X}$, whereas for Data=0.4,
there is a vast and clear gap between actual links and null connections, assuring
perfect reconstruction (Fig.~\ref{fig:TprFpr}(a)). Even with strong measurement noise, e.g.,
$\mathcal{N}(0,0.3^2)$, by increasing Data, full reconstruction can be still accomplished
(Fig.~\ref{fig:TprFpr}(b)). We use two standard indices, true positive rate (TPR) versus false
positive rate (FPR), and Precision versus Recall to measure quantitatively reconstruction
performance~\cite{auroc} (see~\cite{AUC_SM} for more details).
We see that for Data=0.4, both the area under the receiver operating characteristic curve (AUROC)
in TPR vs. FPR (Fig.~\ref{fig:TprFpr}(c)) and the area under the precision-recall curve (AUPR) in
Precision vs. Recall (Fig.~\ref{fig:TprFpr}(d)) equal 1, indicating that links and
null connections can be completely distinguished from each other with a certain threshold.
Because high reconstruction accuracy can always be achieved, we explore the minimum data for
assuring 0.95 AUROC and AUPR simultaneously for different types of dynamics and networks.
As displayed in Table~\ref{tab:simu},
with little measurement noise and a small fraction of inaccessible nodes, only a small amount of
data are required, especially for large networks, e.g., $N=1000$. In the presence of
strong noise and a large fraction of missing nodes, high accuracy can be still
achieved from a relatively larger amount of data. We have also tested our method on
several empirical networks (Table~\ref{tab:real}), finding that only sparse data are
required for full reconstruction as well. These results demonstrate that our general
approach offers robust reconstruction of complex networks from sparse data.

\begin{table}[h]
\caption{Minimum data for achieving at least 0.95 AUROC and AUPR simultaneously for UG, RN and CN
in combination with several real networks. The variables have the same meanings as in Table~\ref{tab:simu}.
See~\cite{SM_empirical} for more details.}
\begin{tabular}{p{1.3cm}<{\centering}p{2.5cm}<{\centering}p{1.3cm}<{\centering}p{1.3cm}<{\centering}
p{1.3cm}<{\centering}}
\hline
\hline
\rule{0pt}{3ex}
& Networks & $N$ & $\langle k\rangle$ & Data\\[0.5ex]
\hline
\rule{0pt}{3ex}
\multirow{3}{*}{UG} & Karate & 34 & 4.6 & 0.69\\[0.3ex]
& Dolphins & 62 & 5.1 & 0.50\\[0.3ex]
& Netscience & 1589 & 3.5 & 0.07\\[0.3ex]
\hline
\rule{0pt}{3ex}
\multirow{3}{*}{RN} & IEEE39BUS & 39 & 2.4 & 0.33\\[0.3ex]
& IEEE118BUS & 118 & 3.0 & 0.23\\[0.3ex]
& IEEE300BUS & 300 & 2.7 & 0.10\\[0.3ex]
\hline
\rule{0pt}{3ex}
\multirow{3}{*}{CN} & Football & 115 & 10.7 & 0.35\\[0.3ex]
& Jazz & 198 & 27.7 & 0.49\\[0.3ex]
& Email & 1133 & 9.6 & 0.10\\[0.5ex]
\hline
\hline
\end{tabular}
\label{tab:real}
\end{table}

In conclusion, we develop a general framework to reconstruct complex networks
with great robustness from sparse data that in general can be much less than network
sizes. The key to our method lies in decomposing the task of reconstructing the
whole network into inferring local connections of nodes individually.
Due to the natural sparsity of complex networks, recovering local structures
from time series can be converted into a sparse signal reconstruction problem
that can be resolved by using the lasso, in which both the error control term and the
L1-norm jointly enable robust reconstruction from sparse data. Insofar as all local structures
are ascertained, the whole network can be assembled by simply matching them.
Our method has been validated by the combinations of three representative dynamical processes
and a variety of model and real networks with noise and inaccessible nodes.
High reconstruction accuracy can be achieved for all cases from relatively
small amounts of data.

It is noteworthy that our reconstruction framework is quite flexible and not limited to the
networked systems considered here. The crucial issue is to find a certain relationship
between local structures and measurable data
to construct the reconstruction form $\mathbf{Y}=\Phi \mathbf{X}$.
Indeed, there is no general manner to establish the reconstruction form for different
networked systems, implying that the application scope of our approach is yet not completely
known. Nevertheless, our method could have broad applications in many fields
due to its sparse data requirement and its advantages in robustness against noise and
missing information. In addition, network reconstruction allows us to infer intrinsic nodal dynamics
from time series by canceling the influence from neighbors~\cite{SM}, although this is beyond
our current scope. Taken together, our approach offers deeper understanding of complex networked
systems from observable data and has potential applications in predicting and
controlling collective dynamics of complex systems, especially when we encounter
explosive growth of data in the information era.


\begin{thebibliography}{99}
%
%
%
%
%
%
%
%
%
%
%
%
%
%
%
%
%
%
%
%
%
%


\bibitem{taming}
A.-L. Barab\'{a}si, Nat. Phys. {\bf 1}, 68 (2005).

\bibitem{boccaletti}
S. Boccaletti, V. Latora, Y. Moreno, M. Chavez and D.-U. Hwang, Phys. Rep. {\bf 424}, 175 (2006).

\bibitem{newman}
M. Newman, {\it Networks: An Introduction} (Oxford University Press, 2010)

\bibitem{hecker}
 M. Hecker, S. Lambeck, S. Toepferb, E. van Someren and R. Guthke, BioSystems {\bf 96}, 86 (2009).

\bibitem{timmereview}
M. Timme and J. Casadiego, J. Phys. A: Math. Theor. {\bf 47}, 343001 (2014).

\bibitem{credit}
G. Caldarelli, A. Chessa, A. Gabrielli, F. Pammolli and M. Puliga, Nat. Phys. {\bf 9}, 125 (2013)

%

%
%
%
%

\bibitem{strogatz}
S. H. Strogatz, Nature {\bf 410}, 268 (2001)


\bibitem{takeover}
A.-L. Barab\'{a}si, Nat. Phys. {\bf 8}, 14 (2011).


\bibitem{raredata}
S. Hempel, A. Koseska, J. Kurths and Z. Nikoloski, Phys. Rev. Lett. {\bf 107}, 054101 (2011).


\bibitem{connectivity}
 M. Timme, Phys. Rev. Lett. {\bf 98}, 224101 (2007).
 \bibitem{topology}
D. Napoletani and T. D. Sauer, Phys. Rev. E. {\bf 77}, 026103 (2008).


 \bibitem{catastrophe}
 W.-X. Wang, R. Yang, Y.-C. Lai, V. Kovanis and C. Grebogi, Phys. Rev. Lett. {\bf 106}, 154101 (2011).

  \bibitem{pg}
 W.-X. Wang, Y.-C. Lai, C. Grebogi and J. Ye, Phys. Rev. X. {\bf 1}, 021021 (2011).

\bibitem{shen}
Z. Shen, W.-X. Wang, Y. Fan, Z. Di and Y.-C. Lai, Nat. Commun. {\bf 5}, 1 (2014).

\bibitem{auroc}
D. Marbach {\sl et al.}, Nat. Methods. {\bf 9}, 796 (2012).


\bibitem{indirect}
B. Barzel and A.-L. Barab\'asi, Nat. Biotechnol. {\bf 31}, 720 (2013).

\bibitem{feizi}
S. Feizi, D. Marbach, M. M\'{e}dard and M Kellis1, Nat. Biotechnol. {\bf 31}, 726 (2013).




\bibitem{hastie}
T. Hastie, R. Tibshirani and J. Friedman, {\it The Elements of
Statistical Learning: Data Mining, Inference, and Prediction, Second Edition} (Springer, New York, 2008)

\bibitem{python}
F. Pedregosa  {\sl et al.}, JMLR. {\bf 12}, 2825 (2011)


\bibitem{CS}
D. L. Donoho, IEEE Trans. Inf. Theory {\bf 52}, 1289 (2006).



\bibitem{altruism}
E. Fehr and U. Fischbacher, Nature. {\bf 425}, 785 (2003).

\bibitem{transportation}
W.-X. Wang and Y.-C. Lai, Phys. Rev. E. {\bf 80}, 036109 (2009)

\bibitem{communication}
M. Welzl, {\it Network Congestion Control: Managing Internet Traffic} (Wiley, New York, 2005).


\bibitem{szolnoki}
A. Szolnoki, M. Perc and G. Szab\'{o}, Phys. Rev. Lett. {\bf 109}, 078701 (2012).

\bibitem{nowak}
M. A. Nowak, K. M. Page and K. Sigmund, Science {\bf 289}, 1773 (2000).

\bibitem{game:review}
G. Szab\'o and G. F\'ath, Phys. Rep. {\bf 446}, 97 (2007).

\bibitem{kuperman}
M. N. Kuperman and S. Risau-Gusman, Eur. Phys. J. B, {\bf 62}, 233 (2008).

\bibitem{er}
P. Erd\"{o}s and A. R\'{e}nyi, Publ. Math. Debrecen. {\bf 6}, 290 (1959).

\bibitem{ws}
D. J. Watts and S. H. Strogatz, Nature (London) {\bf 393}, 440 (1998).

\bibitem{ba}
A.-L. Barab\'{a}si and R. Albert, Science {\bf 286}, 509 (1999).

\bibitem{AUC_SM}
See Supplementary Material [url], which includes Ref.~\cite{naturemethods}.

\bibitem{naturemethods}
D. Marbach, J. C. Costello and R. K\'{u}ffner, {\sl et al}, Nat. Methods, {\bf 9}, 796 (2012).

\bibitem{SM}
See Supplementary Material [url].

\bibitem{SM_empirical}
See Supplementary Material [url], which include Refs.~\cite{karate,dolphins,netscience,ieee39,ieee118,ieee300,football,jazz,email}

\bibitem{karate}
W. W. Zachary, J. Anthropol. Res. {\bf 33}, 452 (1977).

\bibitem{dolphins}
D. Lusseau, {\sl et al},Behav. Ecol. Sociobiol {\bf 54}, 396 (2003).

\bibitem{netscience}
M. E. J. Newman, Phys. Rev. E. {\bf 74} 036104 (2006).

\bibitem{ieee39}
M. A. Pai {\it Energy function analysis for power system stability} (Springer, 1989).

\bibitem{ieee118}
P. M. Mahadev and R. D. Christie, IEEE Trans. Power Syst. {\bf 8}, 1084 (1993).

\bibitem{ieee300}
H. Glatvitsch ,F. Alvarado, IEEE Trans. Power Syst. {\bf 13}, 1013 (1998).

\bibitem{football}
M. Girvan and M. E. J. Newman, Proc. Natl. Acad. Sci. USA {\bf 99}, 7821 (2002).

\bibitem{jazz}
P. Gleiser and L. Danon , Adv. Complex Syst. {\bf 6}, 565 (2003).

\bibitem{email}
R. Guimera, L. Danon, A. Diaz-Guilera, F. Giralt and A. Arenas, Phys. Rev. E , {\bf 68}, 065103, (2003).

\end{thebibliography}
\end{document}